\documentclass[final,3p,times,twocolumn]{elsarticle}

\usepackage{graphicx}
\usepackage{subfig}
\usepackage{dcolumn}
\usepackage{amsmath}
\usepackage{amssymb}
\usepackage{bm}
\usepackage{amssymb}

\usepackage{lineno}

\journal{Computer Physics Communications}

\begin{document}

\begin{frontmatter}


\title{STARlight: A Monte Carlo simulation program for ultra-peripheral collisions of relativistic ions}



\author{Spencer R. Klein}\address{Lawrence Berkeley National Laboratory, Berkeley CA, 94720 USA}
\author{Joakim Nystrand}\address{University of Bergen, Norway}
\author{Janet Seger and Yuri Gorbunov}\address{Creighton University, Omaha, NE 68178 USA}
\author{Joey Butterworth}\address{Rice University, Houston TX, 77251 USA}


\begin{abstract}
Ultra-peripheral collisions (UPCs) have been a significant source of study at RHIC and the LHC.  In these collisions, the two colliding nuclei interact electromagnetically, via two-photon or photonuclear interactions, but not hadronically; they effectively miss each other.  Photonuclear interactions produce vector meson states or more general photonuclear final states, while two-photon interactions can produce lepton or meson pairs, or single mesons. 
 In these interactions, the collision geometry plays a major role.  We present a program, STARlight, that calculates the cross-sections for a variety of UPC final states and also creates, via Monte Carlo simulation, events for use in determining detector efficiency.
\end{abstract}


\end{frontmatter}


\section*{Program summary}

\noindent Program Title: STARlight (v2.2) \hfill\newline
Program obtainable from http://starlight.hepforge.org/  \hfill \newline
No. of lines in distributed program, including test data,etc.: 53,188\hfill\newline
No. of bytes in distributed program, including test data,etc.: 2,965,010\hfill\newline
Programming Language: C++ \hfill\newline
Computer: PCs and workstations \hfill\newline
Operating System: Linux \hfill\newline
RAM: Depends on problem size, but should not challenge modern computers. Around 5 MB for the parameters in sample input files.  \hfill\newline
Classification: 11.2, 17.8, 2.5  \hfill\newline
External Routines: PYTHIA 8.2 and DPMJET 3.0 are needed for some final states. 
\hfill \newline
Nature of problem:  The cross-section for ultra-peripheral collisions is obtained by integrating the photon fluxes in transverse impact parameter space, subject to the requirement (which is also impact parameter dependent) that the colliding nuclei do not interact hadronically.  The program is a two step process.  First, it calculates the cross-sections for the reaction of interest, as a function of $W$ (photon-Pomeron or two-photon center of mass energy), $Y$ (final state rapidity) and $p_T$ (final state transverse momentum).  Second, STARlight generates Monte Carlo events which can be used to determine cross-sections within specific kinematic constraints or for studies of detector efficiencies. The second step includes the decay of any unstable particles produced in the reaction, with appropriate consideration of particle spins and parity.  It outputs these events in ASCII format.  
\hfill\newline
Solution method:   The program generates a two dimensional look-up table of the production cross-section as a function of final state rapidity and mass. The dimensions of the table are selectable, allowing the user to choose the desired accuracy. For certain final states, a second two-dimensional look-up table, giving the transverse momentum distribution, as a function of rapidity, is also used.  
With these look-up tables, the program generates final states.  Particle decays and the final angular distributions are calculated for each event.
\hfill\newline
Restrictions:   The program is focused on ultra-relativistic collisions at Brookhaven's RHIC (Relativistic Heavy Ion Collider) and CERN's LHC (Large Hadron Collider), with final states that are visible in a central detector.  At lower energies (i.e. at the CERN SPS), caution should be exercised because STARlight does not account for the longitudinal momentum transfer to the nucleus; this is larger at low beam energies.  
\hfill\newline
Running time:  The running time depends on the binning used for the look-up tables and the number of events requested.   Table generation is typically less than 15 minutes on a single core, and event generation is quick -- generating 100,000 events takes only a few minutes. Running the provided sample input file to produce 1,000 rho mesons at RHIC energies takes $\sim$7 minutes; running the provided sample input file to produce 1,000 electron-positron pairs at RHIC energies takes $\sim$2 minutes.  
\hfill\newline
References: http://starlight.hepforge.org and references in this article.

\section{Introduction}

Ultra-peripheral collisions (UPCs) \cite{reviews} are an active field of study at the Relativistic Heavy-Ion Collider (RHIC) at Brookhaven National Laboratory \cite{STARfourpion,STARrho,STARee,STARinterference,STAR62GeV,PHENIXjpsi}, at the Fermilab Tevatron  \cite{CDFee} and at the Large Hadron Collider (LHC) \cite{Abelev:2012ba,Abbas:2013oua,TheALICE:2014dwa,Adam:2015gsa,Adam:2015sia,Aaij:2013jxj,Aaij:2014iea,Aaij:2015kea}.  In UPCs, the two nuclei interact electromagnetically rather than hadronically by physically missing each other.  Two-photon and photonuclear interactions can occur. STARlight is capable of simulating a range of two-photon and photonuclear interactions: two-photon production of single mesons and photoproduction of a variety of vector mesons.  It can also interface with the DPMJET III event generator to generate general photonuclear interactions.  Short-lived mesons (including vector mesons) are usually decayed within STARlight and take into account the spin state.  For more complex cases, such as mesons with multiple final states, PYTHIA is used for the decays, but the spin information is lost.

In ultra-peripheral collisions, two types of interactions can be distinguished: purely electromagnetic, usually referred to as two-photon interactions, or photonuclear. 

In a two-photon interaction, the two electromagnetic fields, carried by the two nuclei, collide.  These fields couple to any particles that carry charge, so a wide variety of final states are possible.   For relativistic nuclei, the electric and magnetic fields are perpendicular, and the configuration may be represented as a flux of almost-real photons, following the Weizs\"acker-Williams method.  At an impact parameter $b$ from a relativistic nucleus with charge $Z$ and Lorentz boost $\gamma$, the photon number density is \cite{reviews}
\begin{equation}
N(k,b) = \frac{Z^2 \alpha}{\pi^2} \, \frac{k}{(\hbar c)^2} \, \frac{1}{\gamma^2} \, \big[K_1^2(x) + \frac{1}{\gamma^2}K_0^2(x)\big]
\label{eq:fflux}
\end{equation}
where  $k$ is the photon energy, $x = kb/\gamma\hbar c$,
$\alpha\approx 1/137$ is the electromagnetic fine structure constant and $K_1$ and $K_0$ are modified Bessel functions.  The output  $N$ is the number of photons per unit area, per unit energy. The first term gives the flux of transversely polarized photons; it dominates for relativistic nuclei, and it is the only one considered further here.  As long as the ions are moderately relativistic, the second term is unimportant.

Through these two-photon interactions, pairs of leptons, quarks or electroweak bosons may be produced. If a quark pair is produced, they hadronize into a single meson (with the appropriate spin/parity), a meson pair or a more complex final state. The cross-section depends on the overlap of the two photon fields and on the two-photon coupling to the final state.  STARlight can generate a number of single-meson and lepton pair final states.  

In photonuclear interactions, the most common exclusive final state is a vector meson.
In the vector meson dominance picture, photons fluctuate to $q\overline q$ pairs, which scatter elastically from the target nucleus. For heavy nuclear targets, the amplitudes for scattering from individual nucleons add coherently when the $4-$momentum transfer $q$ is small ($|q|<\hbar/R_A$), leading to a large cross-section.
The treatment of coherence is handled via a Glauber formalism which accounts for multiple interactions within the nucleus \cite{KN99}. 
  
STARlight also simulates photon scattering at higher $p_T$, where the amplitudes from the multiple target nucleons add incoherently.

In photonuclear UPCs, either nucleus can emit the photon, with the other nucleus serving as a target.  These possibilities are indistinguishable.  
Exchanging the photon emitter and target introduces a propagator $\exp{(-i\vec{k}\cdot\vec{b})}$ to account for the position difference.  Here $\vec{k}$ and $\vec{b}$ are the photon and impact parameter vectors \cite{KNinterfere}.    
For 
symmetric collisions (e.g. proton-proton or gold-gold), swapping the nucleus is equivalent to a parity transformation.   Vector mesons are negative parity, so the amplitudes subtract and, for $p_T\rightarrow 0$, vector meson production disappears.  For proton-antiproton collisions, swapping the nuclei is a charge-parity transformation; since vector mesons are CP positive, the two amplitudes add \cite{KNpp}.

One important aspect of ultra-peripheral collisions is that, at least for heavy-ions, multiple interactions are possible between a single ion pair.  This has important consequences; it allows for studies involving correlations among the multiple final state particles.  For example, one can study the azimuthal angular correlations between the neutrons emitted in mutual Coulomb dissociation \cite{Bauretal} or rapidity correlations for states like $\rho^0\rho^0$
\cite{Klusek-Gawenda:2013dka}.

Experimentally, UPCs accompanied by single or multiple Coulomb excitation have proven to be very important.  The nuclear decays produce neutrons which are easily visible in zero degree calorimeters, greatly simplifying triggering. Many experimental results to-date have relied on these neutrons for triggering.  Because of the kinematics of the photon emission and absorption \cite{Bauretal}, to a large degree, each photon acts independently, and the cross-section for multiple reactions is given by
\begin{equation}
\sigma_{123} = \int d^2b P_1(b) P_2(b)...P_n(b)
\label{eq:sigma123}
\end{equation}
where the $P_i(b)$ are the probabilities for the different subreactions, as a function of impact parameter. These probabilities may be substantial; a unitarization procedure corrects for the possibilities of multiple subreactions, such as the production of two $e^+e^-$ pairs. STARlight can calculate cross-sections and generate events for two-photon and photonuclear interactions accompanied by mutual Coulomb excitation.  These are three or four photon interactions: one or two for the photonuclear or two-photon interaction, plus one to excite each nucleus.  It can also calculate the cross-sections for reactions accompanied by mutual excitation to a Giant Dipole Resonance (GDR) interactions; the GDR excitations usually decay by single neutron emission. 

STARlight can simulate the collision of two dissimilar nuclei; both $Z$ and $A$ are selectable by the user. Heavy nuclei are modeled either as hard spheres, or following the Woods-Saxon mass distribution.  For selected nuclei (gold, lead, copper), STARlight uses measured parameters for the nuclear radius \cite{datatable}.
Otherwise, the Woods-Saxon radius is set to $R_A = 1.2\ {\rm fm}\cdot A^{1/3}$.  A skin depth of 0.53 fm is used for all nuclei.  For light nuclei, with $Z\leq 6$, a Gaussian mass distribution is used.  

Each projectile beam energy are defined by the Lorentz factor $\gamma$ set by the user. The input beam energies can be asymmetric for the two projectiles, but internally all calculations are performed in the center of mass frame. Before the output is written, the final state particles are boosted to the laboratory frame specified by the user. 
STARlight is optimized for RHIC and LHC energies, although other energy ranges should be usable. 

STARlight 2.2 should closely reproduce the calculations in the original publications~\cite{KN99,KNinterfere,Lund98}, with the following exceptions. First, for some nuclei, the nuclear radii are set to the measured values, while previously a parameterization assuming $R \propto A^{1/3}$ was always used; this new radius is used exclusively. This reduces the cross-section for gold beams by about 7\%, but has almost no effect for lead beams. Second, the parameterizations of the $\gamma + p \rightarrow J/\psi+p$ and $\gamma + p \rightarrow \Upsilon + p$ cross-sections have been improved, as described below. 

\section{Photon spectra and form factors}

The photon spectrum is calculated in impact parameter space by integrating Eq.~\ref{eq:fflux} over all impact parameters  
under the assumption that the beam projectiles do not interact. 
For photonuclear interactions, the flux is
\begin{equation}
\frac{dN_{\gamma}(k)}{dk} = \int d^2b P_{\rm NOHAD}(\vec{b}) N(k,\vec{b}) 
\end{equation}
and for two-photon interactions, 
\begin{align}
\nonumber\frac{d^2N_{\gamma \gamma} (k_1, k_2)}{dk_1 dk_2} = \\
\int \int  d^2b_1 d^2b_2^2 P_{\rm NOHAD}(|\vec{b_1}-\vec{b_2}|) N(k_1,\vec{b}_1) N(k_2,\vec{b}_2) 
\label{eq:gglum}
\end{align}
Here, $N(k_1,\vec{b}_1)$ is the photon density for photons with energy $k_1$ at position $\vec{b}_1$ from nucleus 1 and $N(k_2,\vec{b}_2)$ is the corresponding density from nucleus 2. The vectors $\vec{b}_1$ and $\vec{b}_2$ have their origins at the center of each nucleus. So, $|\vec{b_1}-\vec{b_2}|$ is the impact parameter between the two nuclei.   The requirement that there be no hadronic interactions reduces the usable luminosity.  The actual cross-section is sensitive to how this non-interaction is defined \cite{Lund98,Jackson90,Hencken95}.  The probability of not having a hadronic interaction at impact parameter $b$, $P_{NOHAD}(\vec{b})$, is implemented in different ways for nucleus-nucleus, proton-nucleus and proton-proton collisions.  

STARlight uses a trick to lower the dimensionality of the integral in Eq. \ref{eq:gglum}.  The integral $\int d^2b_1 \int d^2 b_2$ is replaced with $\int db_1 \int db_2 \int d\theta\ P_{NOHAD} (b)$ where $b=\sqrt{b_1^2+b_2^2-2b_1b_2cos(\theta)}$ \ \cite{Jackson90,Baur90}.

In nucleus-nucleus collisions, the probability to have no hadronic 
interaction at impact parameter $b$ is 
\begin{equation}
P_{\rm NOHAD}(\vec{b}) = e^{-\sigma_{\mathrm{NN}} T_{\mathrm{AA}} (\vec{b})}
\end{equation}
where $T_{\mathrm{AA}} (\vec{b})$ is the nuclear overlap function, 
which is calculated from the nuclear density profiles.
The nuclear densities are assumed to follow a Woods-Saxon distribution, and the nucleon-nucleon interaction cross-sections follow the particle data group parameterization for $pp$ collisions for proton-proton center of mass energies $\sqrt{s}$ above 7 GeV \cite{PDG}:
\begin{equation}
\sigma=(33.73 + 0.2838\ln^2(r)+ 13.67r^{-0.412}-7.77r^{-0.5626}) {\rm mb}
\end{equation}
where $r=s/{\rm 1 GeV}^2$. 
The use of $P_{\rm NOHAD}(b)$ reduces the $\gamma\gamma$ luminosities by up to 20\% compared to the simpler requirement that $b>2R_A$  \cite{Lund98}, because there is still a substantial probability of two nuclei interacting hadronically at impact parameters larger than $2R_A$.

In proton-nucleus collisions, the hadronic interaction probability is also 
calculated from the Glauber model. The probability of having no hadronic interactions is  
\begin{equation}
P_{\rm NOHAD}(\vec{b}) = e^{-\sigma_{\mathrm{NN}} T_{\mathrm{A}} (\vec{b})}
\end{equation}
where $T_{\mathrm{A}} (\vec{b})$ is the nuclear thickness function. For large nuclei, applying this probability function is roughly equivalent to requiring $b > R_A$. This probability is used whether the photon is emitted by the nucleus or the proton. So, the effective photon flux from the proton in a proton-nucleus collision is considerably smaller than if the flux is calculated from just the proton form factor \cite{DZ}. 

In proton-proton collisions the probability of having no hadronic interactions is calculated from 
\begin{equation}
P_{\rm NOHAD}(\vec{b}) = | 1 - \Gamma(s,\vec{b}) |^2 
\end{equation}
where $\Gamma(s,\vec{b})$ is the Fourier transform of the pp elastic scattering amplitude \cite{Frankfurt:2006jp}. This is modeled by an exponential
\begin{equation}
\Gamma(s,\vec{b}) = e^{-b^2/2 b_0}
\end{equation}
with $b_0 = 19.8$~GeV$^{-2}$. This interaction probability corresponds roughly to a cut in minimum impact parameter of $b > 1.4$~fm. 

The nuclear form factor is the convolution of a hard-sphere form factor with a Yukawa potential with a range of 0.7 fm.  This is very close to the the Woods-Saxon distribution, and the form factor may be determined analytically \cite{KN99}.  For a 4-momentum transfer $q$,
\begin{equation}
F(q) = \frac{4\pi\rho_0}{Aq^3} \big[\sin(qR_A)-qR_Acos(qR_A)\big] \big[\frac{1}{1+a^2q^2}\big].
\label{eq:ff}
\end{equation}
Here, $\rho_0$ is the nuclear density. 

For light nuclei, with $Z\leq 6$, a Gaussian form factor is used.  Proton-proton and proton-nucleus collisions are treated separately, as discussed below, with the dipole form factor for the proton \cite{KNpp,DZ}
\begin{equation}
F(q) = \frac{1}{(1+q^2/(0.71 {\rm GeV}^2))^2}.
\end{equation}

\section{Two-photon interactions}

The two photon energies, $k_1$ and $k_2$, are related to the final state invariant mass W and rapidity Y through
\begin{equation}
W = \sqrt{4 k_1 k_2}
\end{equation}
and
\begin{equation}
Y = \frac{1}{2} \ln(\frac{k_1}{k_2}) .
\end{equation}
The two-photon luminosity $d^2N/dk_1 dk_2$ in Eq.~\ref{eq:gglum} can thus be transformed to $d^2N/dWdY$. 
For two-photon interactions, the cross-section to produce a final state $X$ factorizes into a two-photon cross-section and a two-photon luminosity \cite{reviews}:
\begin{equation}
\sigma = \int\!\int \frac{d^2N_{\gamma \gamma}}{dW dY}  \sigma(\gamma_1\gamma_2\rightarrow X) dY dW \; .
\label{eq:sigmatot}
\end{equation}

The limits on $|Y|$ (a symmetric region is assumed) and $W$ in the integrals of Eq. \ref{eq:sigmatot} are set by the user. 

In differential form, Eq.~\ref{eq:sigmatot} becomes 
\begin{equation}
\frac{d^2\sigma}{dW dY} = \frac{d^2N_{\gamma \gamma}}{dW dY}  \sigma(\gamma_1\gamma_2\rightarrow X)  \; .
\label{eq:dsigmadydw}
\end{equation}
The look-up table produced during the first phase of the simulation is based on this equation, which thus provides the basis for the invariant mass and rapidity distributions. 

The final state transverse momentum $p_T$ is the vector sum of the $p_T$ of the two photons.  The $p_T$ distribution of the photons depends on the nuclear form factor \cite{KNinterfere,Baur90,hencken,Vidovic} as defined by 
\begin{equation}
\frac{dN(k,p_T)}{dp_T} = \frac{2F^2 (Q^2= p_T^2) p_T^3}{(2\pi)^2((k/\gamma)^2+p_T^2)^2} .
\label{eq:ggpt}
\end{equation}
Because the $p_T$ distribution depends on the photon energy, it cannot be pre-calculated for every condition.   For simplicity, STARlight determines the photon $p_T$ using rejection sampling. The azimuthal angle in the transverse plane is assumed to be uniformly distributed. 

Three types of two-photon interactions are implemented currently: two-photon production of lepton pairs, single meson production and decay, and the reaction $\gamma\gamma\rightarrow\rho^0\rho^0$, followed by $\rho$ decays.

\subsection{Lepton pairs}

STARlight creates lepton pairs using the equivalent photon approximation (EPA) approach.  This is a lowest order calculation, where the photons are treated as massless.  The kinematic distributions are in generally good agreement with the data, except that the EPA approach finds a lower average $p_T$ than in the data. 
A full quantum electrodynamic calculation, including the photon virtuality, found better agreement with the data \cite{STARee}.   Later calculations showed that the STAR result includes room for an additional contribution to the cross-section, due to higher order corrections \cite{baltzHO,Klein:2004is}.  These corrections are expected, but their effects on the kinematic distributions are not well known.  Still, STARlight lepton pairs are useful for most purposes. 

The cross-section to produce a pair of leptons with lepton mass $m$ and pair invariant mass $W$ is given by the Breit-Wheeler formula \cite{Brodsky} 
\begin{align}
\nonumber \sigma(\gamma\gamma\rightarrow l^+l^-) = \\ 
\frac{4\pi\alpha^2}{W^2} 
\bigg[\big(2+\frac{8m^2}{W^2} - \frac{16m^4}{W^4}\big) \ln{(\frac{W+\sqrt{W^2-4m^2}}{2m})} \\
\nonumber -\sqrt{1-\frac{4m^2}{W^2}}\big(1+\frac{4m^2}{W^2}\big)\bigg].
\end{align}
In the region $W \geq 2m$, the cross-section drops very rapidly with increasing $W$, so it is necessary to use a very large number of bins ($\geq$ 100 bins/GeVc$^{-2}$) in $W$ to accurately evaluate the total cross-section. 

The angular distribution of these lepton pairs is given by \cite{Brodsky}
\begin{equation}
G(\theta) = 2 + 4\big(1-\frac{4m^2}{W^2}\big)
\frac{(1-\frac{4m^2}{W^2})\sin^2(\theta)\cos^2(\theta)+ \frac{4m^2}{W^2}}
{(1-(1-\frac{4m^2}{W^2})\cos^2(\theta))^2} .
\end{equation}
where $\theta$ is the angle between the beam direction and one of the leptons, in the lepton-lepton center of mass frame.  Here, we neglect the effect of the photon $p_T$ on the angular distribution.

All three leptons are treated as stable; $\tau$ (and, if desired, $\mu$) decays must be handled externally.  One weakness of this approach (to be corrected in later versions) is that it loses the effects of the correlations between the lepton spins.

\begin{figure} [h]
\centering
\includegraphics[width=0.5\textwidth]{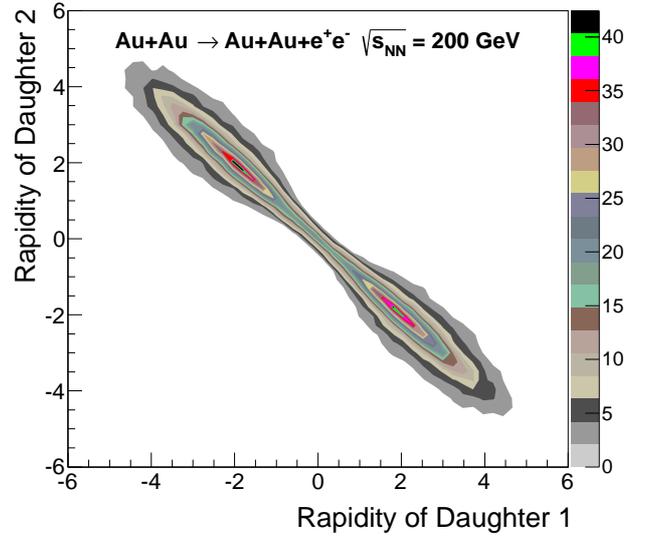}
\caption{Rapidity of one track vs. rapidity of the other for low-mass electron-positron pairs (pair mass between  0.1 and 0.3 GeV/c$^{2}$) produced in $Xn-Xn$ Au+Au collisions at RHIC energies. Pair rapidity is restricted to $|y_{pair}| < 0.1$.}

\label{fig:ee_raprap}
\end{figure}

\begin{figure} [h]
\centering
\includegraphics[width=0.5\textwidth]{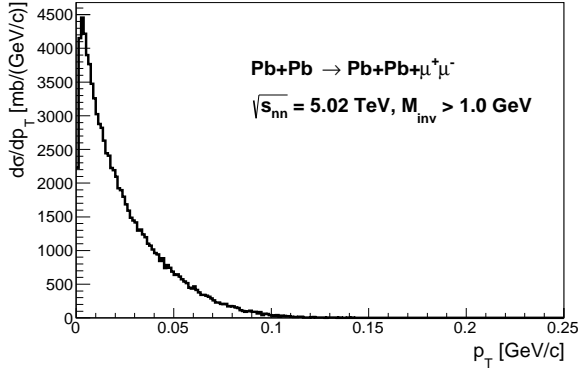}
\caption
{Transverse momentum distribution for muon pairs with pair mass greater than 1 GeV/c$^{2}$ produced in Pb+Pb collisions at LHC energies. }
\label{fig:mumu_pT}
\end{figure}

Examples of the kinematic distributions are shown in Figs. \ref{fig:ee_raprap} and \ref{fig:mumu_pT}.  Figure \ref{fig:ee_raprap} shows the rapidity of each daughter from electron-positron pairs produced in $Xn-Xn$ Au+Au collisions at $\sqrt{s_{NN}}$ $=$ 200 GeV.  Figure \ref{fig:mumu_pT} shows the transverse momentum from $\mu^{-}\mu^{+}$ produced in Pb+Pb collisions at LHC energies.

\subsection{Mesonic final states}

The cross-section to produce a single meson from two photons depends on its two-photon width $\Gamma_{\gamma\gamma}$, mass $M_R$, total width $\Gamma$ and spin $J$ \cite{BreitWheeler}
\begin{eqnarray}
\sigma_{\gamma\gamma}(W) = 8\pi(2J+1)\frac{\Gamma_{\gamma\gamma}\Gamma}{(W^2-M_R^2)^2} \\ \approx 8\pi^2(2J+1)\frac{\Gamma_{\gamma\gamma}}{2M_R^2}\delta(W-M_R) , 
\label{eq:narrow}
\end{eqnarray}
where Eq. \ref{eq:narrow} applies only for narrow resonances.  STARlight can simulate either narrow or wide resonances; in the latter case, the user can select the maximum and minimum $W$.

STARlight simulates the production of a variety of spin 0 and spin 2 mesons, using standard values of $\Gamma_{\gamma\gamma}$.
Short-lived mesons, like the $f_0(980)$ and $f_2(1270)$, are decayed according to their spin and parity.  The mesons are formed by the fusion of two spin 1 photons, with the photon spins either aligned (for spin 2 final states) or antialigned (for spin 0) along the beam axis.  For spin 0, the decays are isotropic, while for spin 2, the meson spin leads to the angular distribution
\begin{equation}
\label{eqn:meson2ang}
\frac {dN}{d\theta} \propto \sin^{5}(\theta).
\end{equation}
This distribution is sampled using rejection sampling.

The f$_2$$'$(1525) is a special case, in that 50\% of the time it decays to $K_sK_s$. These $K_s$'s are written to the output file, and must be decayed by an external program, such as GEANT.

For some two-photon final states, STARlight uses PYTHIA (8.2 or higher) to handle particle decays.  These are the $\eta$, $\eta'$,  $a_2(1320)$ and the $\eta_c$.  The advantage of using PYTHIA is that it can include multiple branching ratios and more complicated final states; the disadvantage is that the spin directional information is lost.

\subsection{$\rho^0\rho^0$}

STARlight can simulate the reaction $\gamma\gamma\rightarrow\rho^0\rho^0$ at threshold ({\it i.e.}, both $\rho^0$ are produced with zero momentum in their center of mass frame) with $\sigma_{\rho^{0}\rho^{0}}$ = 100 nb.  This mode is included to simulate the excess which has been observed in two-photon production, which has been hypothesized to stem from a 4-quark state \cite{Achasov:1991pm}.  

\section{Overview of photonuclear interactions}

STARlight simulates the photoproduction and decay of a variety of vector meson final states: $\rho$, $\omega$, $\phi$, $J/\psi$, $\psi'$, $\Upsilon (1S)$, $\Upsilon (2S)$ and the $\Upsilon (3S)$.  It can also simulate $\rho$+direct $\pi^+\pi^-$ photoproduction, including interference between the two channels. The cross-sections are mostly based on parameterized HERA data for $\gamma p\rightarrow V p$.  It can also simulate general photonuclear interactions
 using DPMJET III. 

The $\rho^0$ and $\omega$ cross-sections, $\sigma(\gamma+p \rightarrow V+p)$, are parameterized as a function of the $\gamma$p center of mass, $W_{\gamma p}$:
\begin{equation}
 \sigma(\gamma+p \rightarrow V+p) = \sigma_{P} \cdot W_{\gamma p}^{\epsilon} + \sigma_{M} \cdot W_{\gamma p}^{\eta} \; .
\end{equation}
The first term is for Pomeron exchange and the second for meson exchange. The other mesons are produced solely via Pomeron exchange, so $\sigma_{M}=0$.

\begin{figure} [h]
\centering
\subfloat{ 
\includegraphics[width=0.24\textwidth]{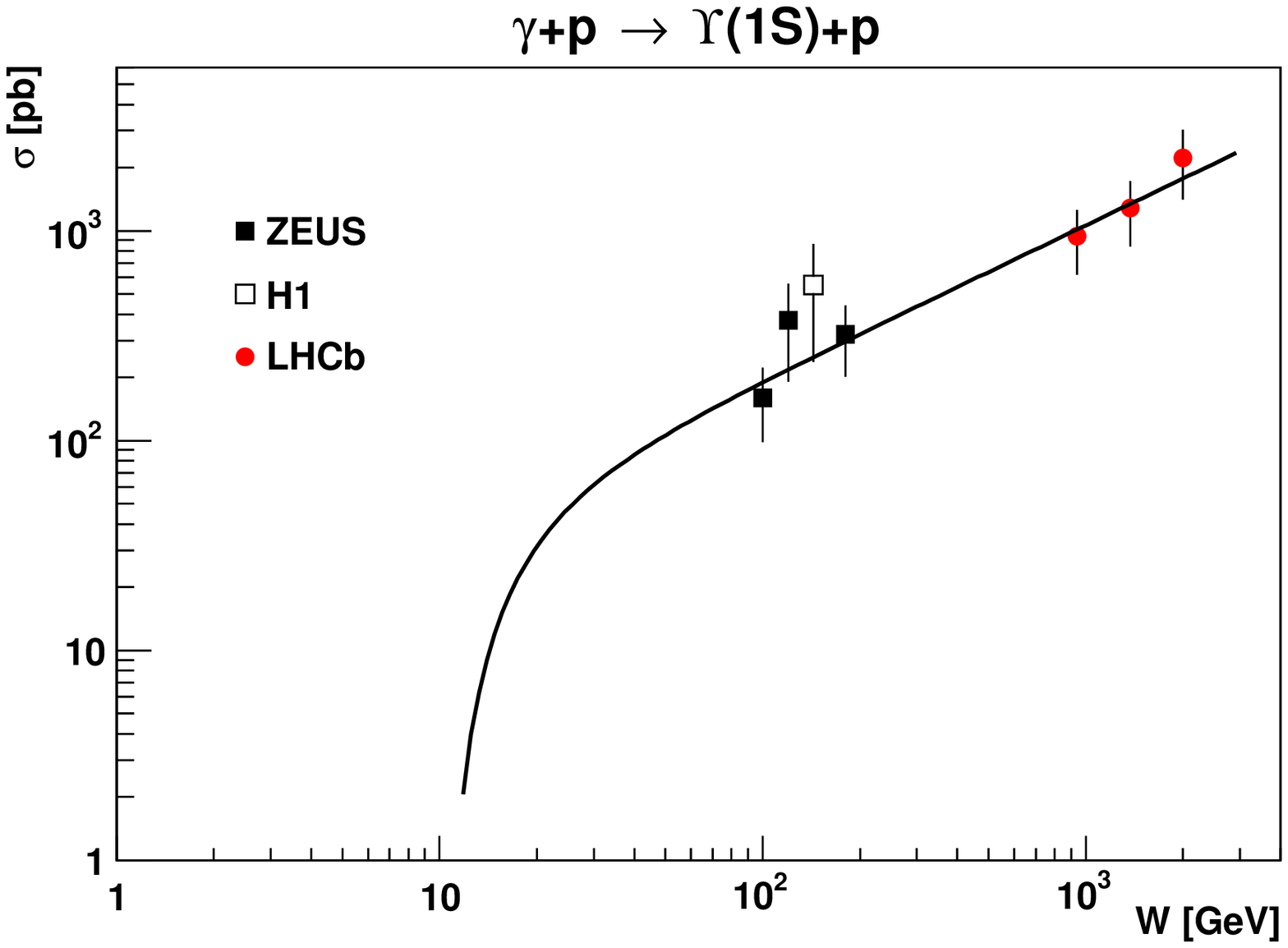}  
\label{fig:UpsilonSlope}}
\subfloat{
\includegraphics[width=0.24\textwidth]{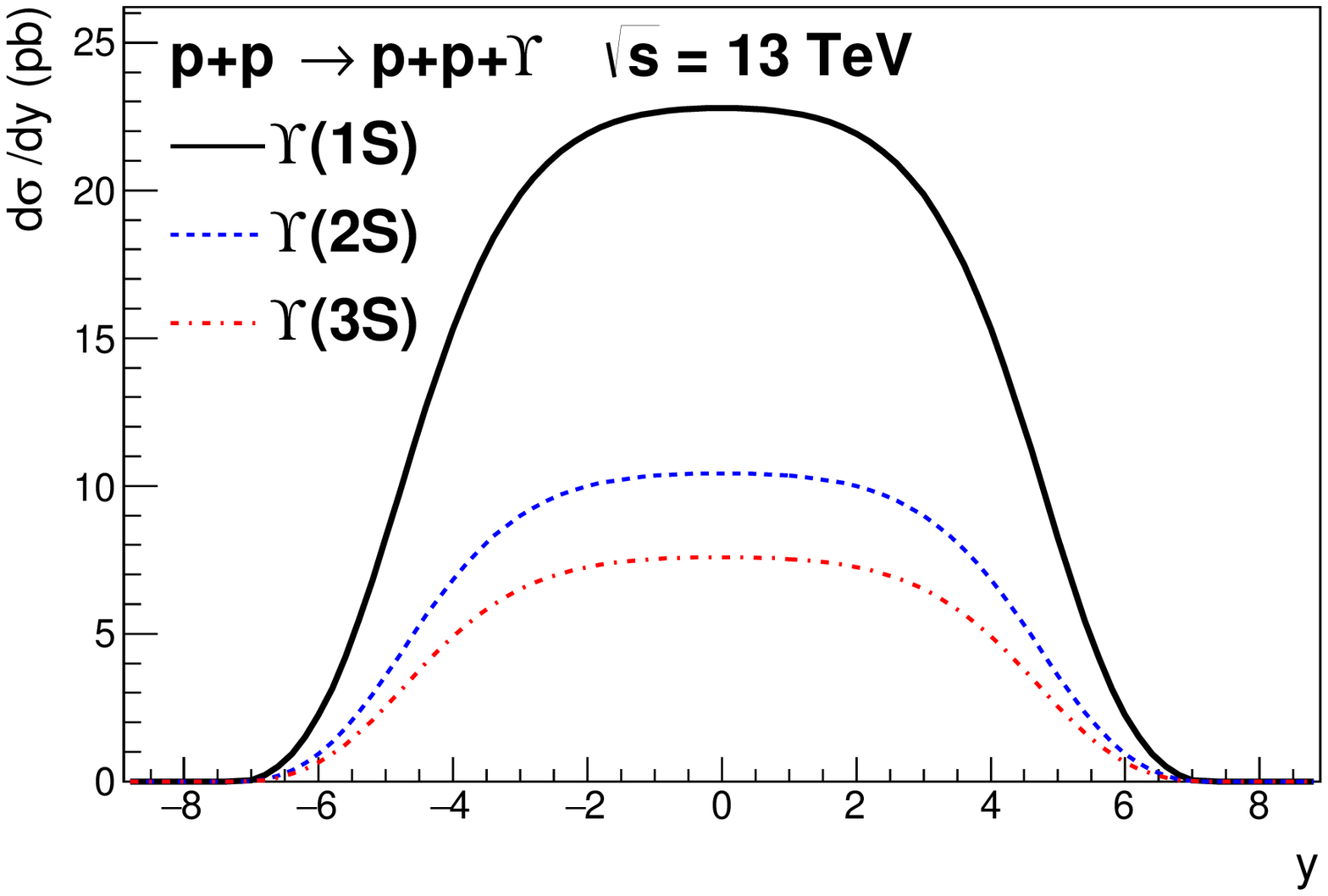}  
\label{fig:Upsilonpp}} \\
\subfloat{
\includegraphics[width=0.24\textwidth]{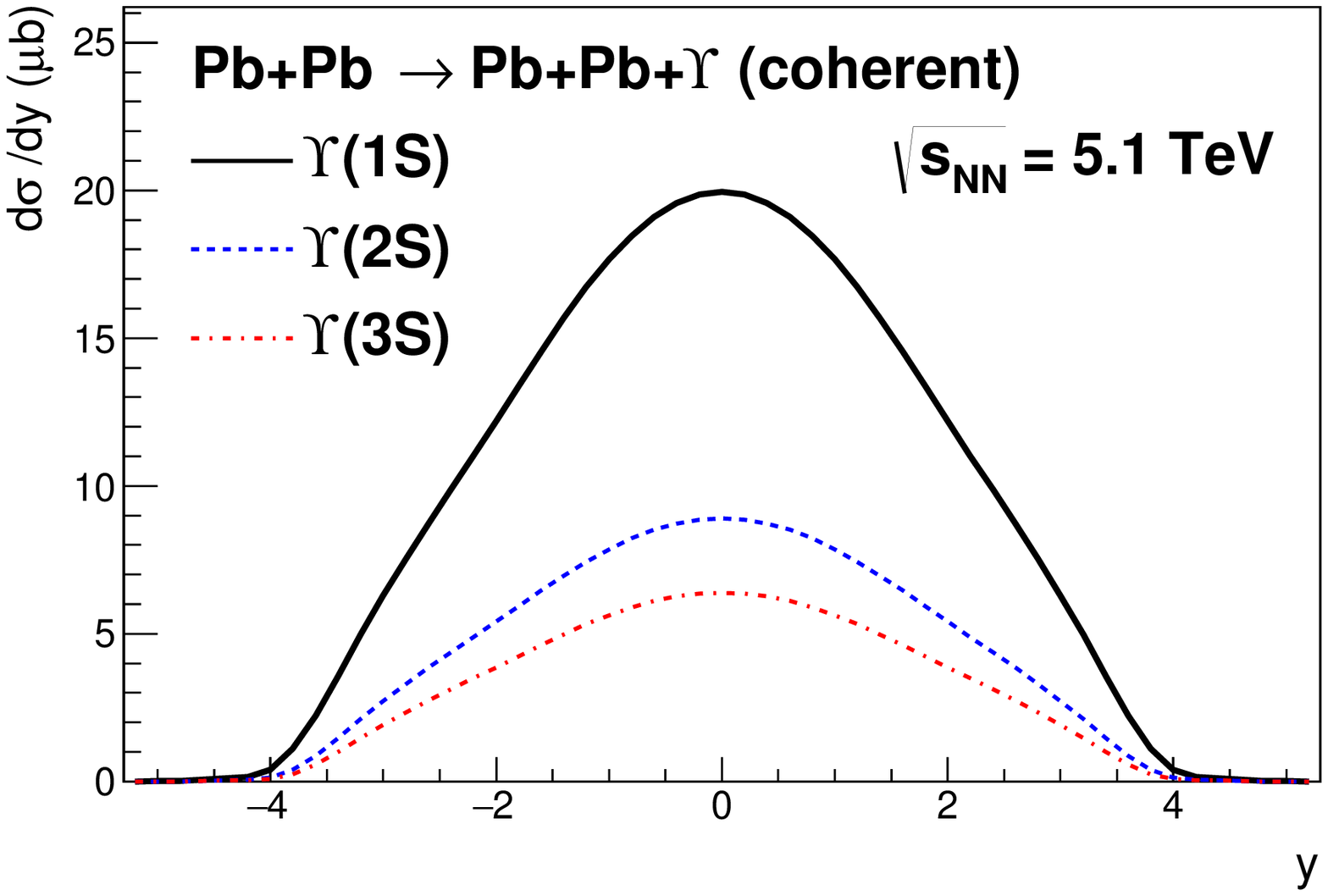}  
\label{fig:UpsilonPbPbCoh}} 
\subfloat{ 
\includegraphics[width=0.24\textwidth]{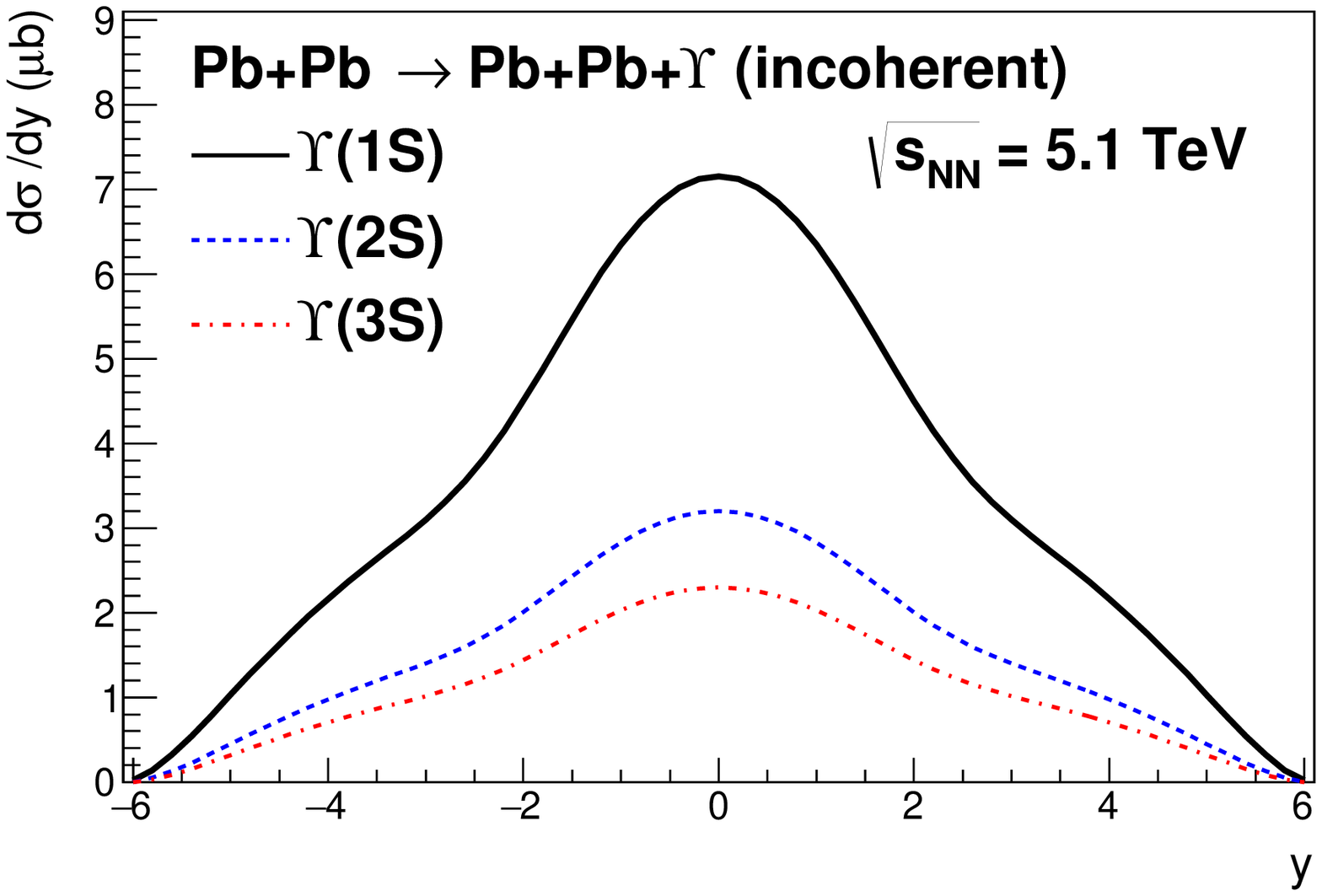}  
\label{fig:UpsilonPbPbIncoh}} 
\caption
{A fit to the available data for the cross-section for Upsilon production as a function of energy in $\gamma$-p interactions is shown in the upper left \cite{ZeusUpsilon,H1Upsilon,Aaij:2015kea}. STARlight uses this fit as the basis for calculating the photonuclear cross-section.  Rapidity distributions for Upsilon mesons produced at LHC energies are shown for $pp$ collisions (upper right), coherent Pb+Pb collisions (lower left) and incoherent Pb+Pb collisions (lower right).}
\label{fig:Upsilon}
\end{figure}

For the $J/\psi$, $\psi'$ and $\Upsilon$ states, the power law is supplemented with a factor that accounts for the near-threshold decrease in the cross-section:
\begin{equation}
\sigma(\gamma+p \rightarrow V+p) = \sigma_{P} \cdot \left[ 1 - \frac{(m_p+m_V)^2}{W_{\gamma p}^2} \right]^2 \cdot W_{\gamma p}^{\epsilon}.
\label{eq:sigmaheavy}
\end{equation}
This is shown in Fig. \ref{fig:UpsilonSlope}.  The cross-sections for $\psi(2S)$ and $\Upsilon(2S,3S)$ states are assumed to have the same shape, with $m_{V}$ replaced by their respective masses. The values of the parameters were obtained from fitting experimental data and are given in Table \ref{tab:vmparameters}.  The rapidity distributions for $\Upsilon (1S)$, $\Upsilon (2S)$ and $\Upsilon (3S)$ produced in $pp$ collisions at $\sqrt{s}$ $=$ 13 TeV are shown in Fig. \ref{fig:Upsilonpp}.  

For most of the particles, the masses and widths are the standard Particle Data Group values.  The exception is the more complex $\rho'$, with considerable evidence for multiple overlapping resonances  \cite{Eidelman}. Because key details about individual states and the interference are poorly known, STARlight uses a single resonance with a mass of 1540 MeV and a width of 570 MeV, chosen to match STAR observations \cite{STARfourpion}. The $\rho'$-photon coupling and nuclear interaction cross-section are arbitrarily taken to be the same as for the $\rho$.  

\begin{table}
\caption{Vector meson parameters for $W_{\gamma p},$ in GeV.  The $\rho$, $\omega$ and $\phi$ parameters are from Ref. \cite{Crittenden}, while the $J/\psi$ values are obtained from a fit of data from \cite{ZeusJpsi,H1Jpsi} to Eq. \ref{eq:sigmaheavy}.  The $\rho'$ characteristics are not well known, so it is arbitrarily given the same parameters as the $\rho$.  The $\Upsilon(1S)$ cross-section uses the parameterization in Eq. \ref{eq:sigmaheavy} and
Fig. \ref{fig:UpsilonSlope}, 
while the $\Upsilon(2S)$ and $\Upsilon(3S)$ cross-sections are scaled from this, based on their couplings to $e^+e^-$ (i.e. $\sigma\propto 1/f_V^2$), per Eq. 10 of Ref. \cite{KN99}.  The
$\psi(2S)$ cross-section is taken to be 0.166 that of the $J/\psi$, following Ref. \cite{PSI2S}. Previous versions of STARlight used older, slightly different values for the $J/\psi$ \cite{KN99}.}
\label{tab:vmparameters}
\begin{tabular}{lrccc}
\hline
Vector Meson& $\sigma_{P}$\ \ \    & $\epsilon$\ \  & $\sigma_{M}$ & $\eta$ \\ \hline
$\rho^0$ \& $\rho'$      & 5.0 $\mu$b\ \ \    & 0.22       & \ 26.0 $\mu$b \  & 1.23    \\
$\omega$       & 0.55 $\mu$b\ \ \   & 0.22       & \ 18.0 $\mu$b\   & 1.92    \\
$\phi$         & 0.34 $\mu$b\ \ \   & 0.22       & --           &  --     \\
$J/\psi$       & 4.06 nb\ \ \       &  0.65      & --           & --      \\
$\psi$(2S)     & 0.674 nb\ \  \     &  0.65      & --           & --      \\
$\Upsilon$(1S) & 6.4 pb\ \ \        &  0.74      & --           & --      \\
$\Upsilon$(2S) & 2.9 pb\ \  \       &  0.74      & --           & --      \\
$\Upsilon$(2S) & 2.1 pb\ \  \       &  0.74      & --           & --      \\	
\hline
\end{tabular}
\end{table}

The vector mesons are decayed assuming that the photon polarization is parallel with the beam axis. 

The same nuclear breakup options as for the two-photon interactions are available. 

\subsection{Coherent vector meson production}

The cross-sections for coherent production on nuclear targets are determined using a classical Glauber calculation \cite{KN99}:
\begin{align}
\nonumber \sigma(AA\rightarrow AAV) =2 \int dk \frac{dN_\gamma(k)}{dk} \sigma(\gamma A\rightarrow VA) \\
= 2 \int_0^\infty dk \frac{dN_\gamma(k)}{dk} \int_{t_{min}}^\infty dt \frac{d\sigma(\gamma A\rightarrow VA)} {dt}\bigg|_{t=0} |F(t)|^2 ,
\label{eq:VMsigma}
\end{align}
where $F(t)$ is the nuclear form factor.  The latter formula is useful because it shows how the $p_T$ distribution of the vector mesons is determined.   

The photon flux $dN_\gamma(k)/dk$ is given by integrating Eq. (\ref{eq:fflux}) over impact parameter space, subject to the condition that the two nuclei do not interact hadronically.  For each impact parameter, the photon flux striking the nucleus is determined and spread evenly over the nucleus.

The calculation of the cross-section can be done assuming narrow resonances, in which case $t_{min} = (M_V^2/4 k \gamma)^2$, or it can be done by convoluting the photon spectrum with the Breit-Wigner shape~\cite{KN99}. This is decided by the user-selected parameter PROD\_MODE. Setting PROD\_MODE=2 gives the narrow resonance approximation and setting  PROD\_MODE=3 uses a wide resonance. The difference is significant only for the $\rho^0$, where the cross-section is reduced by about 5\% in heavy-ion collisions when a wide resonance is used. 

The setting of PROD\_MODE only affects the calculation of the cross section. 
Regardless of the value of PROD\_MODE, the event generation is always done with a Breit-Wigner invariant mass distribution. 

STARlight calculates a two-dimensional look-up table, covering $W$ and $Y$, and then generates $W,Y$ pairs for each event by sampling from the table.  The $W$ and $Y$ range and number of bins are settable, or STARlight can pick a $W$ range to cover the vector meson in question. 

\begin{figure} [h]
\centering
\includegraphics[width=0.5\textwidth]{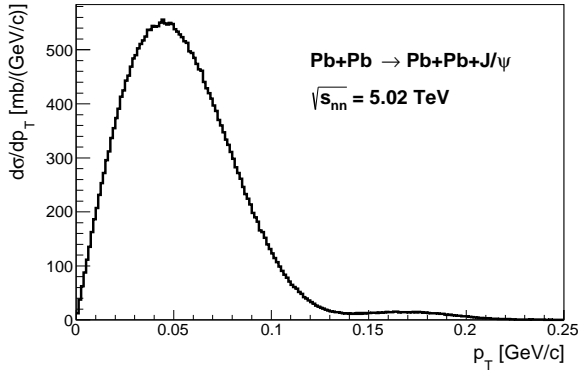}
\caption
{Transverse momentum distribution for J/$\psi$ mesons produced at LHC energies.}
\label{fig:pT}
\end{figure}

There are multiple options for generating the vector meson $p_T$ spectra.  STARlight can generate the $p_T$ spectra assuming that photoproduction on the two nuclei is independent as shown in Fig. \ref{fig:pT} for J/$\psi$ produced in Pb+Pb collisions at LHC energies, or including the interference between the two production sites \cite{KNinterfere}; this is controlled by the user via the INTERFERENCE PARAMETER.  When interference is turned on, it can also generate spectra including interference, but with an altered strength, set by the input parameter IF\_STRENGTH.  To simulate proton anti-proton collisions, where the interference has the opposite sign to proton-proton collisions \cite{KNpp}, one sets IF\_STRENGTH$=-1$.  When interference is used, the $p_T$ spectrum is calculated in another two-dimensional look-up table, in $Y$ and $p_T$; the range and number of $p_T$ bins in this table are settable via the input parameter INT\_PT\_N\_BINS; the number of $Y$ bins is the same as for the cross-section table. 

\subsection{Incoherent vector meson production}

The incoherent photonuclear cross-section is calculated under the assumption 
that it scales the same way as the total inelastic vector meson nucleus 
cross-section. This can be written
\begin{equation}
\frac{\sigma_{\mathrm{inc}}(\gamma + A \rightarrow V+A^*)}{\sigma(\gamma+p \rightarrow V+p)} =
\frac{\sigma_{\mathrm{inel}}(V+A)}{\sigma_{\mathrm{inel}}(V+p)} . 
\end{equation}
Assuming vector meson dominance and using the classical Glauber expression for the total inelastic cross-section, this can also be expressed as 
\begin{equation}
\sigma_{\mathrm{inc}}(\gamma + A \rightarrow V+A) = \frac{4 \pi \alpha}{f_v^2} 
\int \left( 1 - e^{-\sigma_{VN} T(b)} \right) db^2 .
\end{equation}
The term inside the bracket corresponds to the probability of 
having at least one vector meson-nucleon interaction at impact parameter $b$. 
It is thus assumed in the model that one or more vector meson-nucleon interactions 
lead to the emergence of a real vector meson and that it is not destroyed by multiple 
interactions. Other authors have assumed that incoherent production 
occurs only if the vector meson reacts with exactly one nucleon \cite{Lappi:2013am}
or calculated the incoherent cross-section based on event-by-event variations in 
the positions of the nucleons (i.e the variations in the second moment of the amplitudes) \cite{SARTRE}. Other authors have made a direct connection between incoherent interactions and nuclear breakup, leading to neutron emission \cite{Strikman:2005ze}.   These approaches may lead to a reduction in the incoherent cross-section at small $p_T$; this is not present in STARlight.  Figure \ref{fig:UpsilonPbPbIncoh} shows the  $\Upsilon (1S)$, $\Upsilon (2S)$ and $\Upsilon (3S)$ production in incoherent Pb+Pb collisions at $\sqrt{s_{NN}}$ $=$ 5.1 GeV.

The transverse momentum is calculated in the same way as for coherent production, 
but with the nuclear form factor replaced with a nucleon form factor. For the light 
vector mesons, $\rho^0$ and $\omega$, a dipole form factor is used: 
\begin{equation}
F(Q^2) = \frac{1}{(1 + Q^2/Q_0^2)^2} \; .
\end{equation}
This corresponds to a Fourier transform of an exponential matter distribution. The value of $Q_0$ depends on the width of this distribution. The width of the matter distribution is chosen such that the RMS radius is equal to the squared sum of the proton and pion RMS radii, leading to $Q_0^2 = 0.45$~GeV$^2$. 
For the heavier vector mesons, a narrower transverse 
momentum distribution has been observed in $\gamma+p \rightarrow V+p$ data  
and for these an exponential is used for the form factor~\cite{Crittenden}: 
\begin{equation}
F(Q^2) = e^{-b Q^2}
\end{equation} 
with $b=7.0$~GeV$^{-2}$ for $\phi$, $b=4.0$~GeV$^{-2}$ for J/$\psi$ and $\Upsilon$, and 
$b=4.3$~GeV$^{-2}$ for $\psi$(2S).
The distribution used for the light vector mesons is slightly wider than what one gets from an exponential distribution in $Q^2$ with $b \approx 10$~GeV$^{-2}$, as has been found previously~\cite{rho0HERA}, but it gives a good description of data on $\rho^0$ production from ALICE~\cite{Adam:2015gsa}.

\subsection{$\pi^+\pi^-\pi^+\pi^-$ final states}

\begin{figure}  [h]
\centering
\includegraphics[width=0.5\textwidth]{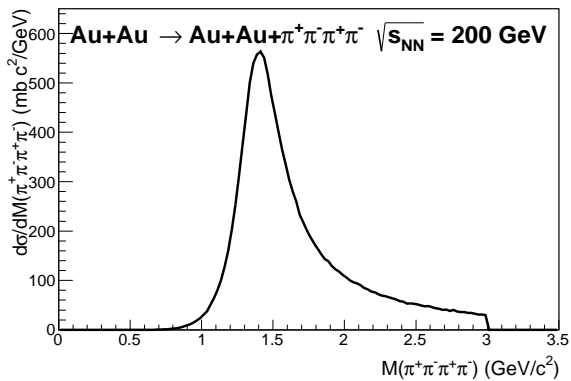}
\caption
{Parent mass distribution for 4-pion final state produced in Au+Au collisions at RHIC energies. The 3 GeV/c$^{2}$ cutoff is due to the choice of $W_{\rm MAX}.$}
\label{fig:4prong_M}
\end{figure}

The structure of the $4\pi$ final state is somewhat complex in that the bulk of the cross-section likely comes from the decay of two or more excited $\rho$ states. In the absence of a detailed model, the 4$\pi$ final state is decayed following a phase space distribution.  Because the final state depends on  the $\rho'-$nucleon cross-section $\sigma_{\rho-N}$, if there are two different vector mesons (with different $\sigma_{\rho-N}$), then the actual $4\pi$ mass spectrum and substructure could depend on the ion species being collided.  The invariant mass distribution of $4\pi$ produced in Au+Au collisions at $\sqrt{s_{NN}}$ $=$ 200 GeV is shown in Fig. \ref{fig:4prong_M}.

\subsection{Interference and $p_T$ spectra}

The vector meson $p_T$ is the vector sum of the photon and Pomeron $p_T$.  The photon $p_T$ is as discussed in the section on two-photon interactions.  The Pomeron $p_T$ is determined by the form factor of the target nucleus, following the distribution in Eq. \ref{eq:VMsigma} with the form factor in Eq. \ref{eq:ff}.  The production amplitude is evenly distributed throughout the nucleus (i.e. absorption is neglected).

Interference between vector meson production on the two nuclear targets is implemented via the cross-section formula:
\begin{equation}
\sigma(y,p_T)= \int d^2b [\sigma(y,p_T) - c \sigma(-y,p_T)]
\label{eq:int}
\end{equation}
where $\sigma(y,p_T)$ gives the $p_T$ spectrum without interference.  Here, $c$ depends on the sign of the combination: c=1 for ion-ion collisions, and $c=-1$ for proton-antiproton collisions ({\it e. g.} at the Tevatron).  In STARlight, $c$ is user settable (input parameter IF\_STRENGTH); it can be any real number. In this approach, Eq. \ref{eq:int} assumes that the production phase is independent of energy.  This should be satisfied for Pomeron exchange, but may fail at low energies where production proceeds via meson exchange.

For incoherent photoproduction, this interference is not important, and is therefore not implemented. 

\subsection{General photonuclear interactions}

STARlight is interfaced with the  DPMJET III \cite{DPMJET} Monte Carlo to simulate general photonuclear interactions $\gamma + A \rightarrow X$. This includes photonuclear jet production and production of soft hadrons. Photon emission from one or both nuclei can be handled, and the photon spectrum is calculated for the two cases as described in \cite{GeneralPhotonuclear}.  Single photon exchange events are characterized by a rapidity gap on the side of the photon-emitting nucleus. The minimum and maximum photon energies (in the rest frame of the target nucleus) must be set by the user through the input parameters MIN\_GAMMA\_ENERGY and MAX\_GAMMA\_ENERGY (in GeV). DPMJET requires that the photon energy is greater than 6 GeV so MIN\_GAMMA\_ENERGY should be set higher than this value. A minimum photon energy of 6 GeV is sufficient to simulate particle production around mid-rapidity ($|\eta| \leq$ 4.5 in Pb+Pb interaction at the LHC), but particle production at very forward rapidities, characterized by low photon energies, cannot be completely described. 

\section{Photonuclear breakup}

Experimentally, it is usually easier to study ultra-peripheral collisions when they are accompanied by photonuclear breakup.  The breakup reaction produces neutrons near the beam rapidity; these neutrons are easily detected in forward calorimeters \cite{STARfourpion,STARrho,STARee}.  

Multiple interactions proceed via multiple independent photons \cite{Bauretal} and are incorporated following Eq. \ref{eq:sigma123}.  Since nuclear breakup depends only on the nucleus-nucleus separation, it can be included without adding any new integrals.  Instead, the two-photon luminosity and the photonuclear cross-section are both multiplied by a probability factor, $P_X(b)$, giving the probability of the desired nuclear breakup at that impact parameter.

STARlight considers two types of nuclear breakup: $Xn$, which is a breakup leading to any number of neutrons, and Giant Dipole Resonance (GDR) excitation, which usually leads to a single neutron emission, $1n$ \cite{Berman:1975tt}. One may require that one or both nuclei break up and the selection is handled by the input parameter BREAKUP\_MODE. 


The cross-section for a nucleus to be excited is \cite{PhotonuclearMCE,TwophotonMCE}
\begin{equation}
P_1 (b) =  \int dk \frac{d^3n(b,k)}{dkd^2b} \sigma_{\gamma A\rightarrow A^*} (k) .
\end{equation}
The subscript '1' shows that this is a non-unitarized probability, and for high enough energy collisions, $P_1(b)$ can exceed one.  The actual probability is given by a unitarization procedure.   
The excitation cross-section $ \sigma_{\gamma A\rightarrow A^*} (k)$ was determined using experimental data \cite{Baltz96}.  Because of this, the excitation code works only for gold or lead beams. For proton-proton or proton-nucleus interactions, only the modes that place no restriction on Coulomb excitation are implemented.     

An interaction with photon energy less than 30 MeV/c (in the target rest frame) is assumed to lead to a Giant-Dipole resonance and, consequently, to a single emitted neutron.  Higher energy photons are assumed to produce higher excitations, with multiple neutrons. 

\section{Program overview}

STARlight runs in two phases.  The first phase calculates cross-sections and other kinematic distributions, and stores them in a series of look-up tables. The second phase uses these look-up tables to rapidly simulate nuclear interactions.  Because of this two-stage approach, it can generate large numbers of events very rapidly.  

\subsection{Look-up table structure \label{sec:loopup}}

The first look-up table to be generated is a two-dimensional table for event mass and rapidity, $d^2N/dYdW$. For photonuclear vector meson production, the shape of the mass distribution is a relativistic Breit-Wigner distribution~\cite{Jackson} with mass and width taken from the PDG. For two-photon production of lepton pairs, the shape of the mass distribution is calculated from the Breit-Wheeler formula (Eq. 18) convoluted with the photon fluxes from each nucleus. For two-photon production of single mesons, the mass distribution is a delta function at the PDG mass.  

Additionally, for photonuclear interactions, a second table stores the $p_T$ distribution, in the form of $d^2 N/dp_T dY$.  In actuality, the $p_T$ depends slightly on $W$, but neglecting this small correlation does not adversely affect the accuracy of STARlight. For $\gamma\gamma$ interactions, the $p_T$ distribution is generated semi-analytically, and no table is needed.

The range of these tables in $Y$ and $W$, and the number of bins for each dimension, are all user settable.  STARlight stores integral tables that are normalized to 1.0.  This simplifies the look-up - for each variable, STARlight selects a random number, finds the appropriate locations in the table that straddle that number, and then linearly interpolates between the two nearest values.    For two-dimensional tables, an additional one-dimensional array holds a collapsed version of the table, just containing $d\sigma/dY$.  Using this, a value of $Y$ is determined.  A second random number is chosen, and the value of $W$ or $p_T$ is found by looking at the two rows in the two-dimensional tables that straddle the chosen value of $Y$.  Then, linear interpolation is performed on the two rows, to determine two values of $W$ corresponding to the $Y$ values for those rows.  Finally, another linear interpolation is performed, in $Y$, to determine the final $W$.  Or course, the same value of $Y$ is used to determine $W$ and $p_T$.

\subsection{Overview of the program structure}

STARlight is coded in C++.  The package contains files in several folders:\\
\textbf{Src} contains the C++ code\\
\textbf{Include} contains the header files that are needed by the C++ code\\
\textbf{Config} contains example input files \\
\textbf{Cmake\_modules} contains cmake macros\\
\textbf{Utils} contains root macros that can be used to convert the output to a root tree, and to produce a simple analysis

There is also a collection of top-level files outside the folder structure including 
\begin{itemize}
\item CMakeLists, a build file
\item Makefile created by CMake
\item Readme contains instructions for installing STARlight, a brief description of the physics incorporated in the code, and a description of input parameters and supported channels
\item starlightDoxyfile.conf gives the settings used by the doxygen documentation system.  doxygen documentation may be created (in the doxygen directory) via "doxygen starlightDoxyfile.conf"
\end{itemize}

The driver program Main.cpp  instantiates, initializes and runs starlightStandalone.cpp.  starlightStandalone.cpp in turns calls methods from STARlight to initialize and produce events of the selected channel and decay them as needed.  The final daughter particles are boosted to the lab frame and written out in ASCII format to the file slight.out.  

The \emph{init} method in the starlight class reads the input parameters from the file slight.in and uses these to determine the beam-beam system and the interaction type (e.g., photon-nucleus vs. two-photon, coherent vs. incoherent).  It checks whether appropriate differential luminosity tables ($dN/dWdY$, with granularity set by input parameters specifying the number of bins in $W$ and $Y$) exist, and calls the appropriate method to create them if they do not.  The differential luminosity tables are stored in the file slight.txt. It instantiates the appropriate event channel and calculates the cross-section. 


Important classes used in the \emph{init} method of STARlight:
\begin{itemize}
\item Nucleus: includes methods to determine nuclear radius, thickness and form factor.
\item Beam: derives from nucleus; includes methods to calculate photon flux, based on impact parameter and photon energy.
\item Beambeamsystem: includes methods to boost to CMS, calculate probability of breakup.
\item photonNucleusCrossSection: base class with methods to calculate cross-section for coherent photon-nucleus interactions.
\item photonNucleusLuminosity: derives from photonNucleusCrossSection, calculates differential luminosity tables for coherent photon-nucleus interactions.
\item incoherentphotonNucleusLuminosity: derives from photonNucleusCrossSection, calculates differential luminosity tables for incoherent photon-nucleus interactions.
\item narrowResonanceCrossSection: derives from photonNucleusCrossSection, includes a method to find the total cross-section for a narrow resonance.
\item wideResonanceCrossSection: derives from photonNucleusCrossSection, includes a method to find the total cross-section for a wide resonance.
\item incoherentVMCrossSection: derives from photonNucleusCrossSection, includes a method to find the total cross-section for incoherent photon-nucleus interactions.
\item twophotonluminosity.cpp: calculates the two photon luminosity table, as a function of $W$ and $Y$.
\end{itemize}

The \emph{run} method in starlightStandalone calls the \emph{produceEvent} method in STARlight.  The \emph{produceEvent} method contains an iterative loop to generate the requested number of events using the method appropriate to the event channel.  All methods derive from the eventChannel class.  Each method randomly picks an energy and rapidity within the selected range, produces the parent system, and decays it where appropriate.  starlightStandalone then boosts the final state particles into the lab frame and writes them into the ASCII file slight.out 

Important classes used in the \emph{produceEvent} method of STARlight:
\begin{itemize}
\item eventChannel: derives from readLuminosity; base class with methods to produce events and keep track of the number attempted and accepted.
\item Gammaavectormeson: derives from eventChannel, includes methods to produce vector mesons from photon-nucleus interactions. 
\item Gammagammaleptonpair: derives from eventChannel, includes methods to create lepton pairs from photon-photon interactions.
\item Gammagammasingle: derives from eventChannel; includes methods to create single-meson events in photon-photon interactions.
\item starlightDpmJet: derives from eventChannel; includes methods to generate photonuclear events using DPMJET. 
\item nBodyPhaseSpaceGen: used for 4-particle final state generation.
\end{itemize}

The \emph{utils} directory includes a few examples which may be useful for some users.  ConvertStarlightAsciiToTree.C converts STARlight output to a ROOT TTree, while Analyze.cxx reads STARlight output and produces some useful histograms. AnalyzeTree.cxx reads the output from ConvertStarlightAsciiToTree and produces similar histograms. 

\section{Description of input data}
STARlight runs are controlled through the slight.in file. The input parameters are listed below, with typical values for LHC Pb+Pb running given in parentheses.   
\begin{enumerate}

\item baseFileName: The name of the output files.  STARlight will copy the input slight.in to baseFileName.in, and produce output files baseFileName.txt and baseFileName.out. (slight)
\item BEAM\_1\_Z: 
Charge of beam one projectile.  (82)
\item BEAM\_1\_A    
Atomic number of beam one projectile. (208)
\item BEAM\_2\_Z:     
Charge of beam two projectile.  (82)
\item BEAM\_2\_A:  	  
Atomic number of beam two projectile. (208)
\item BEAM\_1\_GAMMA: 
Lorentz boost for beam one projectile (p$_z$$>$0). (1470)
\item BEAM\_2\_GAMMA: 
Lorentz boost for beam two projectile (p$_z$$<$0). (1470)
\item W\_MAX: 
Maximum value for the gamma-gamma center of mass energy, W = 4E$_1$E$_2$, in GeV.  Setting W\_MAX = -1 tells STARlight to use the default value specified in inputParameters.cpp (recommended for single meson production). For single mesons, the default W\_MAX is the particle mass plus five times the width. For lepton pairs, the default W\_MAX is given by $2\hbar c\sqrt{\gamma_1\gamma_2 \over{R_1 R_2}}$. (-1) 
\item W\_MIN:     
Minimum value for the gamma-gamma center of mass energy, W = 4E$_1$E$_2$, in GeV.  Setting W\_MIN = -1 tells STARlight to use the default value specified in inputParameters.cpp (recommended for single meson production). The default W\_MIN is the larger of the kinematic limit ({\it e. g.} 2$m_\pi$ for $\rho$ decays) or the particle mass minus five times the width. (-1)
\item W\_N\_BINS: 
Specifies the number of W bins to store in the look-up tables. (40)
\item RAP\_MAX: 
Maximum rapidity of produced particle. (8) 
\item RAP\_N\_BINS: 
Number of rapidity bins used in the cross-section calculation. (80)
\item PROD\_MODE: 
Allows the user to select from a variety of production modes. (2)
\begin{itemize}
\item PROD\_MODE=1:  Two-photon interaction. 
\item PROD\_MODE=2:  Coherent photonuclear vector meson production assuming narrow resonances. 
In pA collisions, this option means that the proton emits the photon and that the gamma-A interaction is coherent. 
\item PROD\_MODE=3:  Coherent photonuclear vector meson production assuming wide resonances. This option should be used for exclusive $\rho^0$ production. 
\item PROD\_MODE=4:  Incoherent photonuclear vector meson production. In pA collisions, this option means that the nucleus emits the photon.
\item PROD\_MODE=5:  Photonuclear one photon exchange uses DPMJET single. 
\item PROD\_MODE=6:  Photonuclear two photon exchange (both nuclei excited) uses DPMJET double.
\item PROD\_MODE=7:  Photonuclear single photon exchange uses DPMJET single proton mode.
\end{itemize}
\item N\_EVENTS:
Number of events to be produced. (1000)
\item PROD\_PID: 
For PROD\_MODE 1 through 4, this selects the channel to be produced, in PDG notation.  Currently supported options are given in Tables 2 and 3. (443013)
\item RND\_SEED:
Seed for random number generator.  (34533)    
\item BREAKUP\_MODE: 
Specifies the way nuclear break-up is handled.  This option only works for lead or gold. In proton-proton or proton-nucleus collisions, options 2 - 7 automatically default to option 5. (5)
\begin{itemize}
\item BREAKUP\_MODE = 1: hard sphere nuclei (no hadronic break-up if impact parameter is greater than the sum of nuclear radii, no restriction on Coulomb break-up).
\item BREAKUP\_MODE =2: requires Coulomb break-up of both nuclei, with no restriction on the number of neutrons emitted by either nucleus ($XnXn$).
\item BREAKUP\_MODE = 3: requires Coulomb break-up of both nuclei, but requires that a single neutron is emitted from each nucleus ($1n1n$).
\item BREAKUP\_MODE = 4: requires Coulomb break-up of neither nucleus. (0n0n)
\item BREAKUP\_MODE = 5: requires that there be no hadronic break up, no restriction on Coulomb break-up (This is similar to option 1, but with the actual hadronic interaction probability).
\item BREAKUP\_MODE = 6: requires Coulomb break up of one or both nuclei, with no restriction on the number of neutrons emitted ($XnXn$ + $0nXn$ + $Xn0n$).
\item BREAKUP\_MODE = 7: requires Coulomb break up of only one nucleus, with no restriction on the number of neutrons emitted ($0nXn$+ $Xn0n$).
\end{itemize}
\item INTERFERENCE: 
Specifies whether interference based on the ambiguity of which nucleus emits the photon is included.  The effect of this interference is only visible at very small transverse momentum.  0 = interference off, 1 = interference on.  (0)
\item IF\_STRENGTH:    
If interference is turned on, specifies the percentage of interference.  The range is -1.0 to 1.0.; 1 is the standard value for ion-ion collisions, while -1.0 is expected for proton-antiproton collisions. (1)
\item INT\_PT\_MAX:   
Used only when the interference option above is turned on.  This specifies the maximum transverse momentum considered, in GeV/c.  (0.24)
\item INT\_PT\_N\_BINS:   
Used only when the interference option above is turned on.  This specifies the number of bins in transverse momentum to use.  (120)
\item INT\_PT\_WIDTH:
Used only when the interference option above is turned on.  This specifies the width of bins in transverse momentum to use.  (0)

\hfill \\
The following parameters are optional. The default value is shown after the parameter name.

\item CUT\_PT: 
Specifies whether the user chooses to place restrictions on the transverse momentum of the decay products. 0= no, 1 = yes. (0)
\item PT\_MIN: 	 
If a transverse momentum cut is applied, this specifies the minimum value produced, in GeV/c. (1.0) 
\item PT\_MAX: 	 
If a transverse momentum cut is applied, this specifies the maximum value produced, in GeV/c. (3.0)
\item CUT\_ETA:  	 
Specifies whether the user chooses to place restrictions on the pseudorapidity of the decay products. 0= no, 1 = yes. (0)
\item ETA\_MIN: 	 
If a pseudorapidity cut is applied, this specifies the minimum value produced. (-10) 
\item ETA\_MAX:	 
If a pseudorapidity cut is applied, this specifies the maximum value produced. (10)
\item XSEC\_METHOD:  
Determines which method is used to calculate the cross-section for $\gamma\gamma$ cross-sections.  XSEC\_METHOD=0 is faster, but works only for symmetric collisions ({\it i. e.} with identical nuclei).  XSEC\_METHOD=1 always works, but is slower. (0)
\newline

The following parameters are used only when interfacing with the PYTHIA and/or DPMJET interfaces:  

\item MIN\_GAMMA\_ENERGY: Allows the user to set the minimum photon energy (in GeV) in the rest frame of the target nucleus. The default is 6.0~GeV and it should never be set below this value since DPMJET was not designed to handle low energy interactions. 
\item MAX\_GAMMA\_ENERGY:  Allows the user to set the maximum photon energy (in GeV) in the rest frame of the target nucleus. The default is 60000.0 GeV.
\item PYTHIA\_PARAMS: Used to supply input parameters to the PYTHIA interface.  This takes a string to pass on semi-colon separated parameters to PYTHIA 6.  {\emph{eg}}: "mstj(1)=0;paru(13)=0.1"  (the default is a blank string " ")
\item PYTHIA\_FULL\_EVENT\_RECORD: Determines whether the full event record from PYTHIA is written to slight.out.  true = yes, false = no (false).  The additional information added is as follows: daughter production vertex (x [mm], y [mm], z [mm], t [mm/c]), mother1, mother2, daughter1, daughter2, PYTHIA particle status code.  The PYTHIA 8 Particle Properties page describes in more detail the properties of mother, daughter, and status code designations.

\end{enumerate}

Currently supported production channels in STARlight are listed in Tables 2 (two-photon channels) and 3 (vector meson channels). 

\begin{table}
\caption{List of two-photon channels in STARlight.}
\begin{tabular}[c]{| l | l |}

\hline
Channel Produced &	PROD\_PID \\ \hline 
e$^+$e$^-$ pair&	11\\ \hline
$\mu^+ \mu^-$ pair&	13\\ \hline
$\tau^+ \tau^-$ pair	& 15\\ \hline
$\rho^0$ pair	& 33\\ \hline
a$_2$(1320) decayed by PYTHIA	&115\\ \hline
$\eta$ decayed by PYTHIA  & 221\\ \hline
f$_2$(1270)$\rightarrow \pi^+\pi^-$ &	225\\ \hline
$\eta$' decayed by PYTHIA	& 331\\ \hline
f$_2$'$(1525)\rightarrow K^+K^- (50\%), K^0\overline K^0 (50\%)$ &	335\\ \hline
$\eta_c$ decayed by PYTHIA	& 441\\ \hline
f$_0$(980)& 	9010221\\ \hline
\end{tabular}
\end{table}

\begin{table}
\caption{List of vector meson channels in STARlight.}
\begin{tabular}[c]{| p{5cm} | l |}

\hline
Channel Produced&	PROD\_PID \\ \hline
$\rho^0 \rightarrow \pi^+ \pi^-$&	113\\ \hline
$\rho^0$ $+$ direct $\pi^+ \pi^-$ production, with interference. The direct $\pi^+ \pi^-$ fraction is from ZEUS data.	& 913\\ \hline
$\omega\rightarrow\pi^+\pi^-$	& 223\\ \hline
$\Phi$	$\rightarrow K^+ K^-$& 333\\ \hline
J/$\psi \rightarrow$ e$^+$e$^-$	& 443011\\ \hline
J/$\psi \rightarrow \mu^+ \mu^-$	& 443013\\ \hline
$\psi$(2S) $\rightarrow$ e$^+$e$^-$		& 444011\\ \hline
$\psi$(2S) $\rightarrow \mu^+ \mu^-$	& 444013\\ \hline
$\Upsilon$(1S) $\rightarrow$ e$^+$e$^-$		& 553011\\ \hline
$\Upsilon$(1S) $\rightarrow \mu^+ \mu^-$	& 553013\\ \hline
$\Upsilon$(2S) $\rightarrow$ e$^+$e$^-$		& 554011\\ \hline
$\Upsilon$(2S) $\rightarrow \mu^+ \mu^-$	& 554013\\ \hline
$\Upsilon$(3S) $\rightarrow$ e$^+$e$^-$		& 555011\\ \hline
$\Upsilon$(3S) $\rightarrow \mu^+ \mu^-$	& 555013\\ \hline
$\rho'\rightarrow\pi^+\pi^-\pi^+\pi^-$      & 999\\ \hline
\end{tabular}
\end{table}

\section{Description of output}
STARlight outputs an ASCII file named slight.out.
For each event, a summary line is printed, with the format

EVENT:  n  ntracks  nvertices ,\\
where n is the event number (starting with 1), ntracks is the number of tracks in the event, and nvertices is the number of vertices in the event (STARlight does not currently produce events with more than one vertex).

This is followed by a line describing the vertex, of the format

VERTEX:  x  y  z  t  nv  nproc  nparent  ndaughters ,\\
where x, y, z and t are the 4-vector components of the vertex location, nv is the vertex number, nproc is a number intended to represent the physical process (always set to 0), nparent is the track number of the parent track (0 for primary vertex) and ndaughters is the number of daughter tracks from this vertex.

This is followed by a series of lines describing each of the daughter tracks emanating from this vertex.  Each track line has the format

TRACK:  GPID  px  py  py nev  ntr  stopv PDGPID ,\\
where GPID is the Geant particle id code, px, py and pz are the three vector components of the track's momentum, nev is the event number, ntr is the number of this track within the vertex (starting with 0), stopv is the vertex number where track ends (0 if track does not terminate within the event), and PDGPID is the Monte Carlo particle ID code endorsed by the Particle Data Group.

\section{Description of test data}
The following parameters were used to produce Fig. 1.\\
\par
\noindent baseFileName =  ee\_RHIC \\
BEAM\_1\_Z = 79 \\   
BEAM\_1\_A = 197 \\  
BEAM\_2\_Z = 79  \\  
BEAM\_2\_A = 197  \\  
BEAM\_1\_GAMMA = 108.4 \\ 
BEAM\_2\_GAMMA = 108.4 \\ 
W\_MAX = 0.3   \\ 
W\_MIN = 0.1   \\ 
W\_N\_BINS = 100   \\ 
RAP\_MAX = 1.3   \\ 
RAP\_N\_BINS = 100  \\ 
PROD\_MODE = 1   \\ 
N\_EVENTS = 1000000  \\ 
PROD\_PID = 11  \\ 
RND\_SEED = 5574533 \\ 
BREAKUP\_MODE = 2   \\ 
INTERFERENCE = 0   \\ 
IF\_STRENGTH = 0.  \\ 
INT\_PT\_MAX = 0.24 \\ 
INT\_PT\_N\_BINS =120  \\ 

Additional input files used to generate the figures in this paper are available in the code distribution, in the \emph{config} folder.

\section{Summary}

STARlight is a versatile C++ code which can simulate a variety of important ultra-peripheral collisions between relativistic heavy nuclei. The output in terms of cross-sections, rapidity and $p_T$ distributions etc. can and has been tested against experimental results. Events generated by STARlight have also been used by several experiments to calculate corrections for acceptance and efficiency; in these calculations, it is important to have an event generator which provide reasonable phase space distributions of the produced particles. STARlight has been found to describe most important features of the data, from both RHIC and the LHC, very well. 

\section{Acknowledgments}

This C++ code was developed from a FORTRAN version. Some of the $\gamma\gamma$ FORTRAN code was written by Evan Scannapieco, and incorporated an earlier basis written by Kai Hencken.   We thank them both profusely.   We thank Boris Grube for useful discussions on photoproduction of the $4\pi$ final state. Kyrre Skjerdal and {\O}ystein Djuvsland added some pieces of code and we thank them for their contribution.

This work was funded by the  U.S.
Department of Energy under contract numbers DE-AC-76SF00098, DE-FG02-96ER40991 and DE-FG02-10ER41666.

\end{document}